\documentclass[aps,pre,reprint,groupedaddress]{revtex4}
\usepackage{graphicx}
\usepackage{dcolumn}
\usepackage{bm}

\newtheorem{theoremc}{Theorem}

\newtheorem{rk}[theoremc]{Remark}

\newcommand\qed{\phantom{\underline{y}}\hfill\hfill$\square$}

\newcommand\p{\partial}

\newcommand\op[1]{\mathop{\rm #1}\nolimits}
\newcommand\po{$\!\!\!{\mbox{\bf.}}$ }

\newcommand{\weg}[1]{}

\begin{document}

\title{Modeling  temporal fluctuations in avalanching systems}

\author{M. Rypdal$^*$ and K. Rypdal$^\dag$  \\~ }
\address{$^*$Department of Mathematics and Statistics, University of Troms{\o}, Norway \\ $^\dag$Department of
Physics and Technology, University of Troms{\o}, Norway }


\begin{abstract}
 We demonstrate how to model the toppling activity in avalanching systems by stochastic differential
equations (SDEs). The theory is developed as a generalization of the classical mean field approach to sandpile dynamics
 by formulating it as a generalization of Itoh's SDE. This equation contains  a fractional Gaussian noise term representing the branching of an avalanche into small active clusters, and a drift term reflecting the tendency for small avalanches to grow and large avalanches to be constricted by the finite system size. If one defines avalanching to take place when the   toppling activity exceeds a certain threshold the stochastic model allows us to compute the avalanche
exponents in the continum limit as functions of the Hurst exponent of the noise. The results are found to agree well with
numerical simulations in the Bak-Tang-Wiesenfeld and Zhang sandpile models. The stochastic model also provides a method for
computing the probability density functions of the fluctuations in the toppling activity itself. We show that the sandpiles do not belong
to the class of phenomena giving rise to universal non-Gaussian probability density functions for the global activity.  Moreover, we demonstrate essential differences between the
fluctuations of total kinetic energy in a two-dimensional turbulence simulation  and the toppling activity
in sandpiles.
\end{abstract}

\maketitle
\section{Introduction\label{sec1}}
The aim of this paper is to present a consistent framework for the modelling of temporal fluctuations, including
definition and computation of avalanche exponents, in sandpile (height) models such as the Bak-Tang-Wiesenfeld (BTW) and Zhang models
\cite{BTW,Zhang}. One of the defining properties of self-organized criticality is that avalanche duration and size are quantities subject to 
scaling, i.e. $p_{\op{dur}}(\tau) \sim \tau^{-\alpha}$ and $p_{\op{size}}(s) \sim s^{-\nu}$
\cite{Jensenbook,Christensenbook}. The calculation of the exponents $\alpha$ and $\nu$ in the thermodynamic limit
$N \to \infty$ ($N^d$ is the number of sites in the $d$-dimensional lattice) has proven to be a difficult task in
dimensions $d=2$ and $d=3$. This is partly due to the lack of simple finite-size scaling in models such as the BTW
model \cite{Christensenbook}. Moreover, it has recently been pointed out \cite{Rypdal} that the difficulty may originate from
the fact that the $\tau$ and $s$ do not scale when defined in the traditional  sense, which is to consider the duration of an avalanche as the time interval where toppling takes place between two successive zeroes in the toppling activity. In this paper we shall denote avalanches defined this way as {\em type-I avalanches}.

The scaling property can be restored, however,  if one defines the duration of an avalanche as the time interval when the toppling activity $x(t)$ exceeds a prescribed threshold $x_{th}$. We shall use the term {\em type-II avalanches} when the start and end of an avalanche is determined via such a threshold criterion. The idea of using a threshold on the toppling activity
in the definition of avalanches was first introduced in \cite{maya}, where it was argued that any avalanche
analysis of real-world activity time series must define avalanches from a threshold, since there is no way to uniquely determine whether a non-negative continuous-valued experimental quantity is actually zero, or just small. For type-II avalanches it can be shown by numerical simulation that the quiet times between avalanches in the BTW model are power-law distributed.  For type-I avalanches the quiet times only depend on the statistics of the driver, which is usually assumed to be  Poisson
distributed. 

The present  work represents the first systematic investigation of  type-II avalanche statistics for the BTW and Zhang
sandpiles. We are particularly interested in  the continuum limit where the system size
$L=1$ is considered fixed and the spatial resolution increases as $N \to \infty$. The power-law statistics of
avalanche observables have cutoffs for large avalanches due to the finiteness of the system, but as we increase
the resolution we see an increasing range of scaling for smaller avalanches. As $N \to \infty$ we keep the threshold fixed relative to the time-averaged activity $\langle x\rangle$, which scales as $\langle x\rangle \sim N^{D_1}$, where $0<D_1<2$. For the BTW sandpile, numerical  simulations yields $D_1\approx 0.86$
\cite{Rypdal}. This means that the number of overcritical sites corresponding to the threshold value
diverges like $x_c\sim N^{D_1}$ in the limit $N \to \infty$.

We model the toppling activity in the continum limit by a stochastic differential equation \cite{Oksendal} for a normalized toppling activity $X(t)$. In its
simplest form this equation is on the form
\begin{equation} \label{SDE1}
dX(t)=\sigma\,\sqrt{X(t)}\,dW(t)\,,
\end{equation}
where $W(t)$ is the Wiener process. Without the factor $\sqrt{X(t)}$ on the right hand side we would simply have  that $X(t)=\sigma W(t)+X_0$ is a Brownian motion with diffusion coefficient $D=\sigma^2/2$. This factor, however, gives rise to a non-uniform ($X$-dependent) diffusion coefficient $D=\sigma^2X(t)/2$, and the stochastic process $X(t)$ will have non-stationary increments. This model can be perceived as a continuous version of the classical mean field
theory of sandpiles \cite{Ivashkevich}. We use its corresponding Fokker-Planck equation  to derive that
$\alpha=2$ and $\nu=3/2$ when avalanches are defined in the type-I sense. This is the same results obtained by
mean field theory \cite{Ivashkevich,Christensenbook,Jensenbook}. For type-II avalanches the effect of the non-uniform diffusion coefficient vanishes for avalanches of durations short compared to that of a system-size avalanche, and for the purposes of
calculating $\alpha$ and $\nu$ we can assume that the the toppling activity is a standard Brownian motion. Solving
the Fokker-Planck equation for Brownian motion (or equivalently using the known distribution of first return times
in Brownian motion) we obtain $\alpha=3/2$ and $\nu=4/3$.

For the modeling of non-trivial sandpile models (BTW and Zhang) the stochastic differential equation takes the
form
\begin{equation}\label{SDE3}
dX(t)=f(X)\,dt+\sigma\,\sqrt{X(t)}\,dW_H(t)\,.
\end{equation}
Two important generalizations of the stochastic model are included here. First, from numerical simulations of sandpiles we find that for small activities $X(t)$ there is an
effective positive drift term. This term is dominant for very small activity, since the diffusion term is negligible for very small $X(t)$ due to the $X$-factor in the diffusion coefficient. The positive drift term is $X$-dependent and quickly decreases as $X$ increases, but
strongly influences the avalanche statistics because it contributes to prevent avalanches from terminating when $X(t)$ approaches zero. We believe that  this effect is responsible for destroying
the scaling of avalanche duration and size (when these are defined in the type-I sense). However, if the drift term
is small compared to the diffusion term for $X>X_{th}$, the drift term will not affect the avalanche statistics if one employs a threshold  $X_{th}$ to
define type-II avalanches. This explains why scaling of size and duration is restored when when type-II avalanches are introduced. Another generalization, which is essential for
the avalanche statistics, is the Hurst exponent $H$ of the noise term. The mean-field approach to sandpiles
implicitly assumes that $H=1/2$. However, this is not the case for the BTW and Zhang models. Actually, analysis of numerical
simulations of the sandpiles show that $H=0.37$ for the BTW model and $H=0.75$ for the Zhang model.

As in the mean-field model, the effect of the non-uniform diffusion coefficient vanishes as the threshold
increases, and hence keeping the threshold $X_{th}$ fixed and increasing $N$ we can, for the purposes of computing
$\alpha$ and $\nu$ for avalanches where $X$ never grows much greater than $X_{th}$, consider the toppling activity as a fractional Browian motion. Using the result of Ding and
Yang \cite{DingYang}, that the first return time in fractional Brownian motion scales like $\sim \tau^{H-2}$ , we
obtain the general results $\alpha=2-H$ and $\nu=2/(1+H)$.

Although the drift term and the  non-uniformity of the diffusion coefficient are not important to calculate the avalanche exponents for type-II avalanches whose duration are short enough not to be limited by the finite system size, they are important on the time scales  where the toppling activity is a stationary process. These are  scales sufficiently long that the toppling  activity of avalanches is limited by the boundaries. The stochastic equation (\ref{SDE3}) is fully equipped to handle these time scales.  A good example of the applicability of these aspects of the stochastic model  is the computation of the probability
density function (PDF) of the temporal fluctuations in the activity signal itself.

For weakly driven sandpiles the PDFs of the fluctuations in the toppling activity are stretched exponentials. This
result is reproduced by simulation of the stochastic model \cite{Rypdal}. For stronger driving the activity
exhibits fluctuations which are more confined around a mean value where the drive and dissipation balance each
other. It has been claimed that the PDFs of the toppling activity in sandpiles are examples of universal
Bramwell-Holdsworth-Pinton (BHP) distributions \cite{Bramwell1,Bramwell2}, a certain class of asymmetric PDFs
commonly seen in complex systems. Our sandpile simulations show that the BHP distributions can only be seen if one fine tunes the
driving rate to a certain value, and for other driving rates the PDFs belong to a much wider class of
distributions. For sufficiently strong drive Gaussian PDFs are observed. These can also be obtained from the stochastic
model if one correctly models the drift term in this parameter range.

The rest of the paper is structured as follows: In Sec.\ \ref{sec2} we explain and derive the stochastic model for the toppling
activity. In Sec.\ \ref{sec3} we compute the avalanche exponents for type-I and type-II avalanches for the mean field
case ($H=1/2$) by solving a Fokker-Planck equation, and for $H \neq 1/2$ we compute the avalanche exponents for type-II avalanches as a function of $H$. The results allow us to predict the avalanche exponents for sandpile models by
computation of the Hurst exponents. These results are then tested against numerical simulations of the
BTW and Zhang models and are shown to agree well. In Sec.\ \ref{sec4} we present results which indicate that type-II avalanches exhibit so-called finite-size scaling, even though type-I avalanches do not.

In Sec.\ \ref{sec5}  we use the stochastic theory to calculate the PDFs of the toppling activity signal, both for
strongly driven sandpiles and in the weak driving limit. The method is finally applied to the fluctuations in
kinetic energy in a two-dimensional turbulence simulation. In this case the process is given by a different kind of stochastic
differential equation:
\begin{equation}
dX(t)=b\,dt+c\,e^{a\,X}\,dW(t)\, . \label{SDE4}
\end{equation}
This equation gives rise to a Fischer-Tippet-Gumbel (FTG) distribution \cite{Gumbel,Coles}, which is  very close to the BHP. The differences between the stochastic models for sandpiles activity and
kinetic energy fluctuations in 2D turbulence may represent an essential distinguishing feature between 2D turbulent dynamics and the kind of avalanching dynamics which are observed in the classical sandpile models.

In Sec.\ \ref{sec6} we summarize and conclude the work.

\section{The stochastic model\label{sec2}}
\noindent
\begin{figure}[t]
\begin{center}
\includegraphics[width=8.5cm]{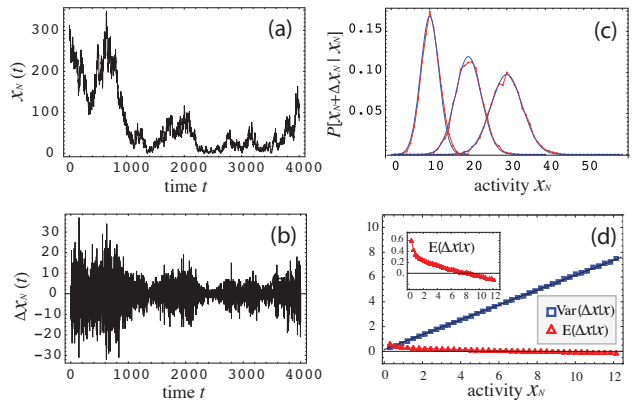}
\caption{{\small a) A realization of the toppling activity $x(k)$ in the BTW sandpile. b) The increments $\Delta
x(k)=x(k+1)-x(k)$ of the trace in (a), showing that $\Delta x(k)$ is large when $x(k)$ is large. c) Conditional
PDFs of $x+\Delta x$ for $x=10,\, 20,\, 30$ respectively. d) The conditional mean and variance of $\Delta x$
versus $x$.}} \label{fig1}
\end{center}
\end{figure}
Let $x(k)$ denote the number of overcritical sites at time step $k$ in a sandpile. A common feature of
(height-type) sandpile models is that the typical size of increments $\Delta x$ is proportional to the square
root of the toppling activity $x$. To be more precise, the conditional probability of an increment $\Delta
x(k)=x(k+1)-x(k)$, given $x=x(k)$, is
\begin{equation} \label{eq1}
P(\Delta x|x)=\frac{1}{\sqrt{2 \pi \sigma^2 x}}\,e^{-\frac{\Delta x^2}{2 \sigma^2 x}}\,.
\end{equation}
In Fig.\ref{fig1} this property is verified for the BTW model ( it holds in the
Zhang model as well). This result can be explained as follows: At a given time $k$ there are $x=x(k)$ overcritical
sites, which we can enumerate $i=1,2,\dots,x$. In the next time step $k \to k+1$ each site $i$ distributes energy to its neighbors, and will usually (always in the BTW model) become subcritical. If none of the neighbors receive sufficient energy to
become overcritical, the contribution to $\Delta x$ from site $i$ is $\xi_i=-1$. If exactly one of the neighbors
become overcritical, then $\xi_i=0$, and so on. For the two-dimensional  models the maximal value of
$\xi_i$ is 4 (3 for the BTW model) since a site maximally can excite 4 neighboring sites. Hence, for this case, we consider $\xi_i$ to
be random variables with realizations in $\{-1,0,1,2,3,4\}$. The randomness originates from the local configuration
in the vicinity of the overcritical site, which for these purposes is considered to be random. In other words, we think of
the configuration on the lattice as a random background.

As an approximation we consider the different realizations of $\xi_i$ (at a fixed time $k$) as independent of each
other. We also consider the distribution of $\xi_i$ to be identical for all overcritical sites and for all times.
In this approximation the total increment $\Delta x$ can be written as a sum of independent, identically distributed, random variables $\Delta
x=\xi_1+\dots+\xi_x$ and by the central limit theorem we have (\ref{eq1}) provided that the local means $\langle
\xi \rangle$ are zero. Then $\sigma^2=\langle \xi^2 \rangle=(-1)^2\,p_{-1}+\dots+4^2\,p_4$, where ${\bf
p}=(p_{-1},\dots,p_4)$ is the probability vector for the local increment processes.

If the local processes at different times $k$ are independent of each other, then $x(k)$ is a
Markov process which satisfies a stochastic difference equation
$$
\Delta x(k)=\sigma \sqrt{x(k)}\,w(k)\,,
$$
where $w(k)$ is a stationary, normalized, and uncorrelated Gaussian process, i.e. $w(k)=W(k+1)-W(k)$, where $W(t)$
is the Wiener process.  Under a rescaling of time $t=k\, \delta t$ and $X(t)=x(k)\, \delta x$ we have
\begin{eqnarray*}
\Delta X(t)&=&X(t+\delta t)-X(t)=\delta x \Big{(} x(k+1)-x(k) \Big{)} \\
&=& \delta x\,\sigma \sqrt{x(k)}\,w(k)= \delta x^{1/2}\,\sigma\, \sqrt{X(t)}\,\Big{(} W(k+1)-W(k) \Big{)} \\
&=& \Big{(} \frac{\delta x}{\delta t} \Big{)}^{1/2}\, \sigma\, \sqrt{X(t)}\,\Big{(} W(t+\delta
t)-W(t) \Big{)}\,.
\end{eqnarray*}
In the last step we used the self-affinity of the Wiener process, $W(t/\delta t){\stackrel{d}{=}} \delta t ^{-1/2}W(t)$.
For $\delta x = \delta t$ we have a well defined model in the limit $\delta t,\delta x \to 0$,  namely the
It{\^ o} stochastic differential equation Eq.~(\ref{SDE1}).

The first generalization of this model is obtained if we relax the requirement that the local increment processes
$\xi_i(k)$ at time $k$ are independent of the local increment processes $\xi_j(k')$, $j=1,\dots,x(k')$ at previous
times $k'<k$. In this case we need to model memory effects in the stationary Gaussian process
$$
w(k)=\frac{\Delta x(k)}{\sigma \sqrt{x(k)}}\,.
$$
From the power spectrum or the  variogram of the activity signal from numerical simulation of the  BTW and Zhang models we find that $w(k)$ can be
accurately modelled as a colored noise characterized by a Hurst exponent $H$. That is, $w(k)=W_H(k+1)-W_H(k)$,
with $W_H$ being a normalized (diffusion coefficient $=1$) fractional Brownian motion (fBm).
If we perform the rescaling  $t=k\, \delta t$ and $X(t)=x(k)\, \delta x$ in the case $H \neq 1/2$ we obtain
$$
\Delta X(t)=\Big{(}\frac{\delta x}{\delta t^{2H}} \Big{)}^{1/2}\, \sigma\, \sqrt{X(t)}\,\Big{(}
W_H(t+\delta t)-W_H(t) \Big{)}\,,
$$
and by requiring that $\delta x = \delta t^h$, with $h=2H$,  we obtain the stochastic differential equation
\begin{equation}
dX(t)=\sigma \sqrt{X(t)}\,dW_H(t)\,. \label{SDE2}
\end{equation}
If we assume  that the the stochastic process $X(t)$ is self-affine with self-affinity exponent $h$, i.e. $X(st)\stackrel{d}{=}s^h X(t)$, it is easy to verify that Eq.~(\ref{SDE2}) is invariant with respect to the transformation $t\rightarrow st$ if $h=2H$. Thus, the exponent $h=2H$ is the self-affinity exponent of the process $X(t)$, where $H$ is the Hurst exponent determining
the color of the noise process driving the stochastic differential equation. Observe that the reason why $h\neq H$ is the non-stationarity of the increment process due to the the factor $\sqrt{X(t)}$ in Eq.~(\ref{SDE2}). Note also that the case $H=1/2$
corresponds to $h=1$. 

The self-affinity of $X(t)$ described by Eq.~(\ref{SDE2}) implies that there is no upper bound on the fluctuations on increasing time scales, i.e. Eq.~(\ref{SDE2}) describes the activity of an infinite sandpile where the activity is never influenced by the system boundaries. From a physical viewpoint, however, it is more interesting to consider the dynamics of a finite sandpile in the continuum (thermodynamic)  limit, and for this purpose it is natural to let the scaling factor $\delta x$ depend on $N$ such that $X_N=x_N\delta x(N)$ is bounded in the limit $N\rightarrow \infty$, for instance such that $\lim_{N\rightarrow \infty}\max (X_N)=1$. Such a bound on $X(t)$ can be obtained by the introduction of a drift term $f(X)\, dt$ leaving the stochastic equation in the form of Eq.~(\ref{SDE3}),
where $f(X)$ is negative for large $X$. The form of $f(X)$ can be found from sandpile simulations by computing the conditional mean $E(\Delta x|x)$ of the increments, and is shown in Fig.\ref{fig1}d. It appears that $f(X)$ is a decreasing function,  positive for small $X$ and negative for large $X$, and $f(X)=0$ for a characteristic activity $X_c\sim 1$. Without the stochastic term the drift term establishes $X_c$ as a stable fixed point for the dynamics. 

Since $f(X=0)>0$ solutions of Eq.~(\ref{SDE3}) with initial condition $X(0)>0$ exist and are positive for all $t>0$. This means that while Eq.~(\ref{SDE2}) has solutions for which $X(t)=0$ after a finite time (avalanches terminate), avalanches described by  Eq.~(\ref{SDE3}) will never terminate in the meaning $X(t)=0$ (type-I avalanches). This signifies that type-I termination never occurs in the the continuum limit. If sandpiles of increasing $N$ are simulated, and the type-I avalanche durations are computed in the rescaled coordinates, the durations generally grow without bounds for increasing $N$. The reason is that the effective threshold for type-I termination in a  discrete sandpile is $x_N=1$, but in rescaled coordinates this threshold  $X_N=x_N\, \delta x(N)$ goes to zero as $N\rightarrow \infty$. As this rescaled threshold vanishes the duration in rescaled time goes to infinity. Since the type-I termination is a discreteness effect, the resulting PDFs of avalanche durations depends on $N$ (system discreteness) and are not power-laws. As we shall demonstrate later, the introduction of activity thresholds which are defined in the rescaled coordinates, and hence remain finite in the continuum limit, will give rise to PDFs of durations of type-II avalanches which converge to a specific power-law in this limit.

Eq.~(\ref{SDE3}) remains valid also for sandpiles which are driven by continuous feeding of sand during avalanches. The drive adds a positive contribution to the drift function $f(X)$ for $X<X_c$, but a negative contribution for $X>X_c$, because for large activites $f(X)$ mainly accounts for the increased   boundary losses. The result is a  steeper $f(X)$, which tends to confine the activity closer to the fixed point $X_c$. On the  other hand the increased drive also increases the diffusion coefficient (by increasing $\sigma$) due to a larger number of new active clusters initiated per unit time. The net effect is a positive shift of $X_c$ and that the
fluctuations in $X$ are confined to a smaller region around $X_c$. For sufficiently strong drive the range of
variation in $X(t)$ becomes so small that the diffusion coefficient does not vary much. It is nevertheless important to
model it correctly in order to calculate the PDFs of the fluctuations in toppling activity.

\section{Calculation of avalanche exponents}\label{sec3}
We denote the stochastic model Eq.~(\ref{SDE1}) (which is Eq.~(\ref{SDE3}) with $H=1/2$ and $f(X)=0$) the {\em mean field model} of sandpiles. This is because
its underlying assumptions and the results derived from it coincide with what is known as the mean field solution
of sandpiles in the literature \cite{Ivashkevich}. Eq.~(\ref{SDE1}), together with its corresponding Fokker-Plack formulation can
be used to calculate the avalanche exponents $\alpha$ and $\nu$. The idea is that each avalanche corresponds to a
realization $X(t)$ with some initial condition $X(0)=X_0$ ($X_0 \ll X_{max}$). The avalanche propagates until the
realization $X(t)$ terminates at $t=t_1$ in the meaning that $X(t)>0$ for $t<t_1$ and  $X(t_1)=0$. Calculating the ratio of surviving realizations at
different times $t$ in an ensemble will provide information about the distribution of avalanche durations. The avalanche
size distribution can then be obtained by using a general relationship between the self-affinity exponent $h=2H=1$
and the duration statistics.

The scenario outlined above can be mathematically formulated as follows: Let $P(X,t)=\op{Pr}[\mbox{$X(t)$ exists
and } X(t)=X]$. Then the the probability $\rho(t)$ that an avalanche still runs after a time $t$ (we call
it the survival function) is given by
\begin{equation}
\rho(t)=\int_0^\infty P(X,t)\,dX\,, \label{duration}
\end{equation}
and the probability density function for durations is $p_{\op{dur}}(\tau)=-\rho'(\tau)$.
The density $P(X,t)$ can be calculated by solving the Fokker-Planck equation
\begin{equation} \label{FokkerPlanck}
\frac{\p P}{\p t}=\frac{\sigma^2}{2}\,\frac{\p^2}{\p X^2}(X\,P)
\end{equation}
on $X\in [0,\infty)$,  $t\in [0,\infty)$, subject to an absorbing boundary condition $\lim_{X \to 0} X P(X,t)=0$ and an initial condition
$P(X,0)=\delta(X-X_0)$.

To correctly incorporate the absorbing boundary condition we let $U=XP$, and solve the corresponding Fokker-Planck
equation for $U$ with boundary condition $U(0)=0$ to get
$$
P(X,t)=\int_0^\infty G(X,Y,t)\,P(Y,0)\,dY\,,
$$
where
$$
G(X,Y,t)=\frac{1}{2}\,\sqrt{\frac{Y}{X}}\,\int_0^\infty
J_1(s\,\sqrt{Y})\,J_1(s\,\sqrt{X})\, \exp{\left(-\frac{\sigma^2\,s^2}{8}\,t \right)}\,s\,ds\,.
$$

\begin{rk} \po
The most elegant way to obtain the solution is to use the integral transform pair
\begin{eqnarray*}
\hat{F}(s)&=&\frac{1}{2}\,\int_0^\infty
F(X)\,J_1(s\,\sqrt{X})\,\sqrt{X}\,dX \\
F(X)&=&\frac{1}{\sqrt{X}}\,\int_0^\infty \hat{F}(s)\,J_1(s\,\sqrt{X})\,s\,ds\,.
\end{eqnarray*}
Taking the transform of the Fokker-Planck equation we obtain the ODEs
$$
\frac{\p \hat{P}}{\p t}(s,t)=-\frac{\sigma^2\,s^2}{4}\,\hat{P}(s,t)\,,
$$
which we can solve and take the inverse transform. It is also easy to solve the Fokker-Plack equation for $U=XP$
by separation of variables.
\end{rk}

If $P(X,0)=\delta(X-X_0)$ the solution of the absorbing boundary problem is $P(X,t)=G(X,X_0,t)$. Moreover
\begin{equation}
\lim_{X \to 0} P(X,t)=\frac{\sqrt{X_0}}{4}\,\int_0^\infty
s^2\,J_1(s\,\sqrt{X_0})\,\exp{\left(-\frac{\sigma^2\,s^2}{8}\,t \right)}\,ds= \frac{4\,
X_0}{\sigma^4 t^2}\,\exp{\left(-\frac{2\,X_0}{\sigma^2\,t}\right) }\,. \label{P0}
\end{equation}
From Eqs.~(\ref{duration}), (\ref{FokkerPlanck}), (\ref{P0}), and the absorbing boundary condition at $X=0$ we find that
$$
\frac{d \rho}{dt}=-\frac{\sigma^2}{2}\,\lim_{X \to 0} P(X,t)=- \frac{2\,
X_0}{\sigma^2 t^2}\, \exp{\left(-\frac{2\,X_0}{\sigma^2\,t}\right)}\,.
$$
Hence the PDF for avalanche durations $\tau$ is
$$
p_{\op{dur}}(\tau)=\frac{2\, X_0}{\sigma ^2 \tau^2}\,\exp{\left(-\frac{2\,X_0}{\sigma ^2\,\tau}\right)}\,,
$$
and for $\tau>>2X_0/\sigma^2$ we have $p_{\op{dur}}(\tau) \sim \tau^{-\alpha}$, with $\alpha=2$.

This result crucially depends on the correct formulation of the Fokker-Planck equation. If the stochastic process
$X(t)$ were a classical Brownan motion, the Fokker-Planck equation would have the form of the standard heat
equation, and by performing the analogous calculations for this equation we get $p_{\op{dur}}(\tau) \sim
\tau^{-3/2}$. The same scaling relation ($\sim \tau^{3/2}$) is obtained if one (incorrectly) employs the Stratonovich
formulation of the Fokker-Planck equation rather than the Itoh form \cite{Oksendal}.

From the derivation of the stochastic model we have seen that the process $X(t)$ has a self-affinity exponent
$h=2H$ (for $H=1/2$ we have $h=1$). This means that $X(t)$ disperses like $\sim t^h$. For long avalanches this
implies that the size of the avalanche scales as
\begin{equation} \label{sizedur}
s=\int_0^\tau X(t)\,dt \sim \int^\tau t^h\,dt \sim \tau^{h+1}\,. \label{s versus tau}
\end{equation}
This property is easy to check directly by studying the relation between duration and sizes of avalanches in
sandpiles. We find that the relation holds for large avalanches, whereas for very small avalanches $s \sim
\tau^1$. If the survival function scales as $\rho(\tau) \sim \tau^{-\delta}=\tau^{-\alpha+1}$, then using
(\ref{sizedur}) together with $p_{\op{size}}(s)\,ds=p_{\op{dur}}(\tau)\,d\tau$ yields that if $p_{\op{size}}(s)
\sim s^{-\nu}$, then
\begin{equation} \label{rel}
\nu=\frac{1+h+\delta}{1+h}=\frac{\alpha+h}{1+h}\,.
\end{equation}
In the case $H=1/2$ ($h=1$) and $\alpha=2$ this gives $\nu=3/2$.
\noindent
\begin{figure}[t]
\begin{center}
\includegraphics[width=12.5cm]{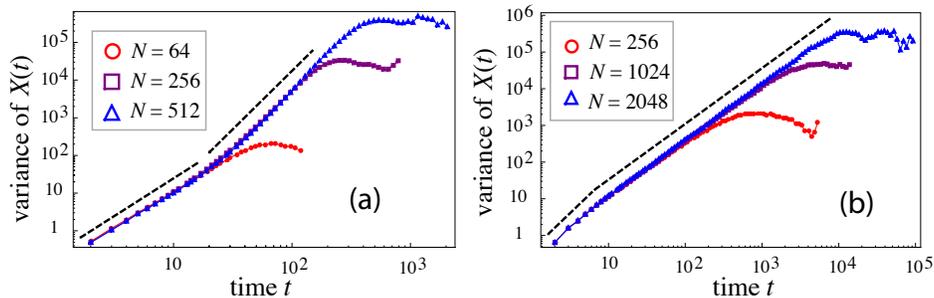}
\caption{{\small a) Determining $h=2H$ in the Zhang model through the calculation of the variance of the activity
$X(k)$ in avalanches still running at time $t$. On shorter time scales there is a Hurst exponent $H=0.5$ (the
first dashed line has slope $2h=4H=2$) on longer time scales the Hurst exponent is $H=0.75$ (the second dashed
line has slope $2h=4H=3$). b)  Determining $h=2H$ in the BTW model by the same method as in (a). In this case we
have $H=0.37$.}} \label{fig4}
\end{center}
\end{figure}

\noindent
\begin{figure}[t]
\begin{center}
\includegraphics[width=12.5cm]{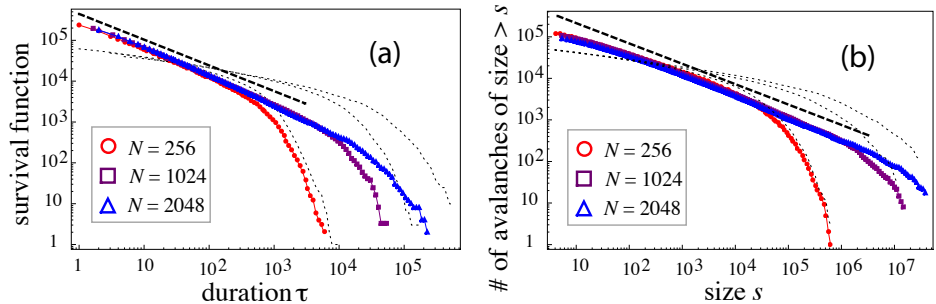}
\caption{{\small a) The survival function of avalanches in the BTW model computed with thresholds $\langle
X \rangle/3$ (type-II) and without thresholds (type-I, the dotted curves). The dashed line corresponds to $\alpha=2-H=1.63$. b)
The probability density function of avalanches with size $>s$ in the BTW model computed with thresholds
$\langle X \rangle/3$ and without thresholds (the dotted curves). The dashed line corresponds to
$\nu=2/(1+H)=1.49$.}} \label{fig5}
\end{center}
\end{figure}
\noindent
\begin{figure}[t]
\begin{center}
\includegraphics[width=12.5cm]{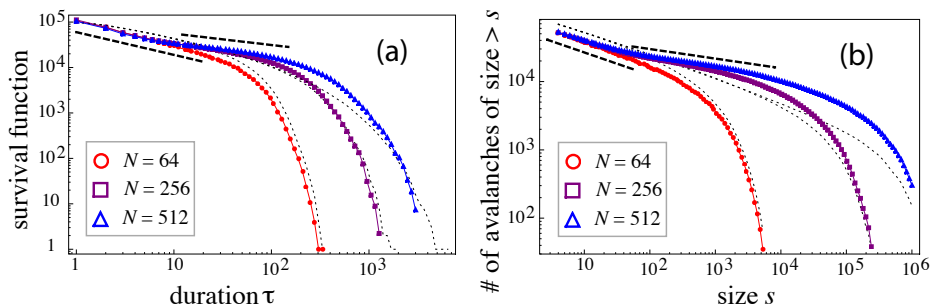}
\caption{{\small a) The probability of having avalanches of duration  $>\tau$ in the Zhang model computed with thresholds
$\langle X \rangle/3$ and without thresholds (the dotted curves). The first dashed line corresponds to
$\alpha=2-H=1.50$ obtained with $H=0.50$, and the second dashed line corresponds to $\alpha=2-H=1.25$ obtained
with $H=0.75$. b) The probability of having avalanches with size $>s$ in the Zhang model computed
with thresholds $\langle X \rangle/3$ and without thresholds (the dotted curves). The first dashed line
corresponds to $\nu=2/(1+H)=1.33$ obtained with $H=0.50$, and the second dashed line corresponds to
$\nu=2/(1+H)=1.14$ obtained with $H=0.75$.}} \label{fig6}
\end{center}
\end{figure}
This mean-field solution is known to be quite correct for the random neighbor sandpile model
\cite{Randomneighbor}, but numerical simulation shows that is fails for type-I as well as type-II avalanches in the BTW and Zhang models. The computation of these exponents are shown for these models in Figures \ref{fig4}, \ref{fig5}, and \ref{fig6}. The computation of $h$ presented in Fig.~\ref{fig4}  for type-I avalanches yield $H=h/2=0.37$ for the BTW model, and  $H=0.75$ for the Zhang model. This invalidates the Fokker-Planck formulation, which can be strictly justified only for a white noise source term ($H=1/2$). This is one reason for the failure of the mean-field approach, but there  are also others.

 For avalanche duration and size type-I  avalanches do not yield good power-law PDFs. The reason for this was discussed in the previous section: letting avalanches terminate when $x_N(t)=0$ in a sandpile with $N^d$ sites corresponds to using an effective threshold for termination in the rescaled coordinates $X_N(t)$ which goes to zero  as $N\rightarrow \infty$. The termination process depends on the discreteness of the system and one cannot expect convergence to scale-invariant behavior in the continuum limit. 

For type-II avalanches Figs.~\ref{fig5} and \ref{fig6} show good scaling for duration and size in BTW and Zhang models, but the scaling exponents differ from those of the mean-field approach. The discrepancy is partly due to the fact that the sandpile models have $H\neq 1/2$, but is also related to the observation that the introduction of a finite termination threshold $X_{th}$ modifies the exponent $h$ as it appears in Eqs.~(\ref{sizedur}) and (\ref{rel}). In the following  we  shall demonstrate that it is possible to obtain analytical results for type-II avalanches from the stochastic model which are in agreement with the corresponding sandpile simulations.

First we observe that omission of the drift term will only have effect on avalanches which are so large that they are strongly limited by boundary dissipation. Next, we notice that the effect of the positive drift term for small activities is eliminated for type-II avalanches if $X_{th}>X_c$. In other words, it is only the cut-offs of the power-law PDFs due to finite system size which are lost by this omission. Thus, by considering a threshold $X_{th}$  which is not much smaller than the mean activity $\langle X\rangle$, and considering only avalanches which are sufficiently small not  to be influenced by the boundary dissipation,  we have $X(t) \sim X_{th}$ and thus can justify the substitution $X(t)\rightarrow X_{th}(t)$ on the right hand side in Eq.~(\ref{SDE2}). This equation  then reduces to the equation for an fBm with Hurst exponent $H$. For an fBm the duration of avalanches is  given by the return time statistics for   $W_H(t)$, which is known to scale like $\tau^{H-2}$ \cite{DingYang}, i.e. we have 
\begin{equation}
\alpha=2-H\, . \label{alfa}
\end{equation}
Now we need to address a slightly subtle point: the exponent $h$, as defined in Eq.~(\ref{sizedur}), is $h=2H$ for times  $\tau$ so large that $X(\tau)\gg X(0)$. Only on such time scales will the effect of the factor $\sqrt{X(t)}$ in the stochastic term show up in the scaling.  However, this makes sense only for type-I avalanches where we can choose $X(0)\ll \langle X\rangle $. For type-II avalanches we have $X(0)\approx X_{th}$, and it is more natural to consider the opposite limit where $\tau$ is so small that $\hat{X}(\tau)\equiv X(\tau)-X_{th}\ll  X(0)\approx X_{th}$. On these time scales the activity measured relative to the threshold level scales like an fBm with Hurst exponent $H$, i.e. $\hat{X}(t)\sim t^{H}$, because $X(t)\approx X_{th}$.  The avalanche size of type-II avalanches defined as $\hat{s}(\tau)\equiv \int_0^{\tau}\hat{X}(t)\, dt$ then scale as $\sim \tau^{1+H}$. Thus, for type-II avalanches  we have $h=H$, and hence Eq.~(\ref{rel}) for this case reduces to
\begin{equation}
\nu=\frac{2}{1+H}\,. \label{nu}
\end{equation}
In the mean field limit $H=1/2$ these exponents reduce to those given in the right hand column in Table \ref{tab1}. These results obtained analytically by approximating the coefficient on right hand side in the stochastic equation by its threshold value has been verified by numerical solutions of the full equation. 
\begin{table}[h]
\caption{Exponents in mean-field solutions ($H=1/2$) of the stochastic equation with zero threshold (type-I) and large threshold (type-II).} \label{tab1}
\begin{center}
\begin{tabular}{||c|c|c||} \hline
~~ & type-I avalanches& type-II avalanches) \\ \hline \hline $H$ & $1/2$ & $1/2$ \\ \hline $h$ & $1$ &
$1/2$ \\ \hline $\alpha$ & $2$ & $3/2$ \\ \hline $\nu$ & $3/2$ & $5/4$ \\ \hline \hline
\end{tabular}
\end{center}
\label{default}
\end{table}%

Eqs.~(\ref{alfa}) and (\ref{nu}) show that the calculations of $\alpha$ and $\nu$ reduces to determining the Hurst exponent of
the normalized increment process
$$
w(k)=\frac{\Delta x(k)}{\sqrt{x(k)}}\,,
$$
which can easily be constructed from the toppling activity signal. The corresponding Hurst exponent of the motion
$$
W(k)=\sum_{i=0}^k w(k)
$$
can be calculated using standard techniques such as taking the power spectrum or constructing variograms.
Alternatively we can find $H$ by computing  the variance of $X(t)$ in
an ensemble of realizations starting with small initial values of $X$. This variance scales like $\langle X^2(t)\rangle \sim t^{2h}=t^{4H}$. Fig.\ref{fig4}
shows this computation for the BTW and Zhang models. Since we find $H=0.37$ for the BTW model the type-II scaling exponents are $\alpha=1.63$ and $\nu=1.49$. This is verified by numerical calculation of
$\alpha$ and $\nu$ using a threshold $\langle x \rangle/3$, as shown in Fig.\ref{fig5}.

In the Zhang model the process $w(k)$ is more complicated. Fig.\ref{fig4} shows that $W(k)$ has a Hurst exponent
$H=0.5$ on short time scales and a different Hurst exponent $H=0.75$ on longer timescales. This means that short
avalanches should satisfy the mean-field solution $\alpha=1.5$ and $\nu=1.33$, whereas longer avalanches should
have exponents $\alpha=1.25$ and $\nu=1.14$.  All of these predictions are verified by the direct calculation of
$\alpha$ and $\nu$ as shown in Fig.\ref{fig5}.

For increasing $N$ the long-avalanche scaling dominates an increasing portion of the graph, so in the continuum limit the mean-field solution only prevails at infinitely small scales in rescaled coordinates. It is however a nice verification of our method to see that the relation between the
avalanche exponents and the Hurst exponent correctly predicts the avalanche exponents on short time scales as
well.

These results on the BTW and Zhang models are summarized in table \ref{tab2}.
\begin{table}[t]
\caption{{\small Exponents for type-II avalanches computed from $H=0.37$ in the BTW model, and $H=0.50$ (short
time scales) and $H=0.75$ (long time scales) in the Zhang model.}} \label{tab2}
\begin{center}
\begin{tabular}{||c|c|c|c||} \hline
~ & BTW & Zhang & Zhang \\
~ & ~ & {\small (short times)} & {\small(long times)} \\ \hline \hline $H$ & 0.37 & 0.50 & 0.75 \\
\hline $\alpha$ & 1.63 & 1.50 & 1.25 \\ \hline $\nu$ & 1.49 & 1.33 & 1.14 \\ \hline \hline
\end{tabular}
\end{center}
\label{default}
\end{table}%

\begin{rk} \po
The numerical simulations of the Zhang model are run using the standard toppling rule $z_i \to 0$ and $z_j \to
z_i+z_i/4$ if $z_i$ is overcritical and $j$ is a nearest neighbor of $i$. Whenever the configuration has no
overcritical sites a random site $i$ is chosen with respect to uniform probability and a mass $\epsilon$ is added to this
site: $z_i=z_i+\epsilon$. In the simulations presented in this paper we use $\epsilon=0.1$. In the strongly driven
Zhang models presented in Sec.~\ref{sec5} the feeding times (which can now be during avalanches) are Poisson
distributed and we have used $\epsilon=0.25$.

For the BTW model we have used the standard toppling rule $z_i \to z_i-4$ and $z_j=z_j+1$ through out the paper.
As usual a mass $\epsilon=1$ is added to a random site whenever the configuration is stable.
\end{rk}

\section{Finite-size scaling for type-II avalanches}\label{sec4}
\noindent
\begin{figure}[h]
\begin{center}
\includegraphics[width=8.5cm]{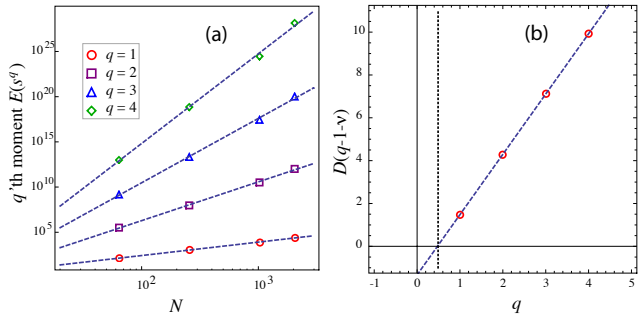}
\caption{{\small a) The moments $\langle s^q \rangle$ plotted as functions of $N$ for the size distribution
of the BTW model ith respect to  a threshold $\langle X \rangle/3$. b) The shape of the structure function $\zeta(q)$. The
slope of the line is $D=2.81$ and the intersection with the first axis is $\nu-1=0.49$. This value is indicated
by the dotted vertical line.}} \label{fig12}
\end{center}
\end{figure}
A systematic technique for the determination of the avalanche exponents $\alpha$ and $\nu$ is the so called moment
analysis \cite{Christensenbook}. We illustrate how this works for the size distribution of the BTW model. The
analysis confirms our result $\nu=0.49$ for the BTW model for type-II avalanches.

Let $p_{\op{size}}(s;N)$ be the PDF of avalanche size $s$ in the two-dimensional BTW model with $N^2$ sites and a threshold $\langle x
\rangle/3$. Let us assume finite-size scaling (FSS). This means that for $N,s \gg 1$ we have
\begin{equation} \label{FSS}
p_{\op{size}}(s;N) \propto s^{-\nu}\,G(s/N^D)\,,
\end{equation}
for some exponent $D>0$, where $G(r)$ falls off quickly for $r>1$. From Eq.~(\ref{FSS} ) we observe  that
$$
\langle s^q \rangle \propto N^{D(1+q-\nu)}\,\int_{1/N^D}^\infty r^{q-\nu}\,G(r)\,dr\,.
$$
The integral tends to a constant as $N \to \infty$ so we have $\langle s^q \rangle \sim N^{D(1+q-\nu)}$. Thus, if we plot $\langle s^q \rangle$ versus $N$ in a log-log plot the slope of a fitted straight line will give us an estimate of the exponent
 $\zeta(q)=D(1+q-\nu)$
for $q=1,2,3,\dots$. Fig.\ref{fig12}a shows the computation of moments of $s$ as a function of $N$ for type-II avalanches obtained from simulation of the BTW model,  and
Fig.\ref{fig12}b shows the exponent $\zeta(q)$ versus $q$. We find that $\zeta(q) \sim q^{2.81}$, hence that
$D=2.81$. Moreover, when plotted in a log-log plot, the intersection of $\zeta(q)$ with the first axis corresponds
to the value $\nu-1$. Hence, based on our predictions this intersection should be in the point $0.49$. This value
is plotted as a dotted horizontal line in Fig.\ref{fig12}b, confirming our result with good accuracy.

\begin{figure}[h]
\begin{center}
\includegraphics[width=8.5cm]{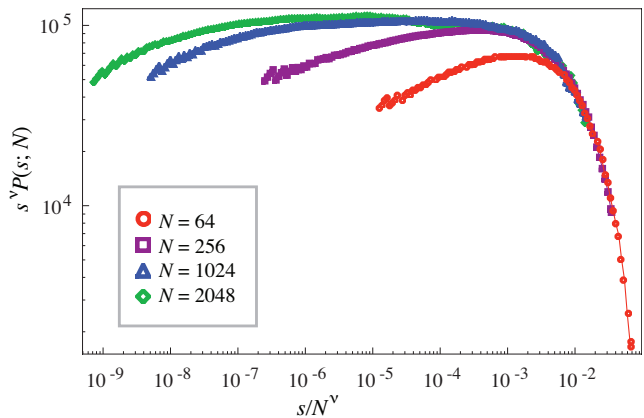}
\caption{{\small Attempted data collapse for the avalanches in the BTW model with respect to a threshold $\langle X
\rangle/3$. We have plotted $s^{\nu-1} \int_s^\infty p_{\op{size}}(s')\,ds'$ as a function of $s/L^D$ with
$D=2.81$ for $N=64,256,1024,2048$.}} \label{fig13}
\end{center}
\end{figure}

Strictly, this method requires that we have data collapse when re-scaling the avalanche sizes by $s/N^D$. The
shape of the scaling function $G(r)$ can be seen by plotting $s^\nu p_{\op{size}}(s;N)$ versus $s/N^D$. This is
shown in Fig.~\ref{fig13}. The  data collapse for the available system sizes is not perfect,  but it is better than for type-I avalanches. In fact, it seems that we might have convergence to a $N$-independent scaling function as $N \to \infty$, and that the lack of data collapse observed here is simply due to the fact that we are not able to
simulate sufficiently large sandpiles. Thus the general lack of scaling for for type-I avalanches, including the finite-size scaling, seems to be restored for type-II avalanches.
\noindent

\section{The PDFs of the toppling activity} \label{sec5}
\noindent
\begin{figure}[t]
\begin{center}
\includegraphics[width=8.5cm]{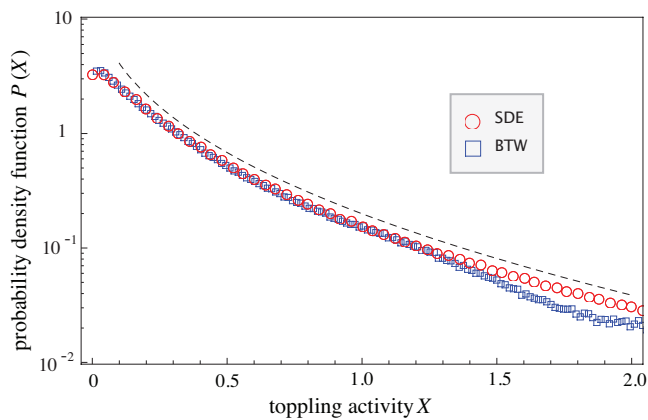}
\caption{{\small The PDF of the fluctuations in the toppling activity of the slowly driven BTW sandpile
together with the corresponding PDF produced from the stochastic model. The dashed line is a fitted stretched
exponential.}} \label{fig7}
\end{center}
\end{figure}
When calculating the PDFs of the toppling activity signal we have to distinguish between the weakly and strongly
driven sandpiles. For the weakly driven sandpiles the toppling activity covers a large range and the high-activity tail of the PDF decays like a
stretched exponential. This property can be reproduced by simulations of Eq.~(\ref{SDE3}) where $f(X)$ 
decays exponentially  to zero as $X$ increases. Fig.\ref{fig7} compares the PDF obtained from such a simulation
of the stochastic differential equation (run with $H=0.37$) with the PDF of the weakly driven BTW model. The
stochastic model is run by initiating new avalanches (realizations) whenever the previous avalanche terminates,
thus representing the classical slow drive of the sandpile model. Since the Hurst exponents of the Zhang and BTW
models are different from $1/2$, the application of the Fokker-Planck equation can not be used to calculate the
shape of the PDFs, and thus we have to rely on numerical simulations.

It is also interesting to study the PDFs of toppling activity in strongly driven sandpiles. In particular in the
light of the recent claims that this toppling activity belongs to a class of fluctuating global quantities with a universal non-gaussian shape \cite{Bramwell1,Bramwell2}. These PDFs are reminiscent of distributions
derived from the extreme value theory of statistics \cite{Gumbel,Coles}, which deals with a sequence of $n$
identically distributed, independent random variables ${y_1,\ldots,y_n}$. If the tail of the PDF of these variables decays
faster than any power-law, the PDF of the $m$'th largest value drawn from each of $M$ realizations of
this sequence converge to the Gumbel class of stable distributions in the limit $M\rightarrow \infty$:
\begin{equation}
G_k(y)=K (e^{-\xi(y-s)} e^{-e^{-\xi(y-s)}})^m. \label{Gumbel}
\end{equation}
The distribution for the largest value in each realization ($m=1$) is often called the Fischer-Tippet-Gumbel
distribution (FTG).  The FTG with zero mean and unit variance requires $K=\xi=\pi/\sqrt{6}$ and $s=\gamma
\sqrt{6}/\pi$, where $\gamma\approx 0.58$ is the so-called Euler constant.

A distribution obtained from the spin-wave approximation to the 2D XY model for equilibrium crititical
fluctuations in a finite-size magnetized system is the so-called Bramwell-Holdsworth-Pinton (BHP) distribution
\cite{Bramwell2}, which corresponds to Eq.\ (\ref{Gumbel}) with $k$ having the non-integer value $m=\pi /2\approx
1.57$.  With $K=2.16$, $\alpha=1.58$, $\xi=0.93$ and $s=0.37$ this distribution has zero mean and unit variance.
The difference between the normalized  FTG and BHP distributions is rather small, and it is difficult to
distinguish between the two based in experimental and numerical data.

Analysis of our simulations show the claim that the toppling activity in strongly driven sandpiles has a PDF similar to  the BHP or FTG distributions is wrong, unless the driving rate of the system is fine tuned to some
particular value. Actually, the normalized PDFs of the toppling activity is only insensitive to variation of  the driving rate in the limits of weak and strong drive. In the limit of weak drive the PDFs are
stretched exponentials as shown in Fig.\ref{fig7} and in the limit of strong drive the PDFs are close to Gaussian.
\noindent
\begin{figure}[t]
\begin{center}
\includegraphics[width=8.5cm]{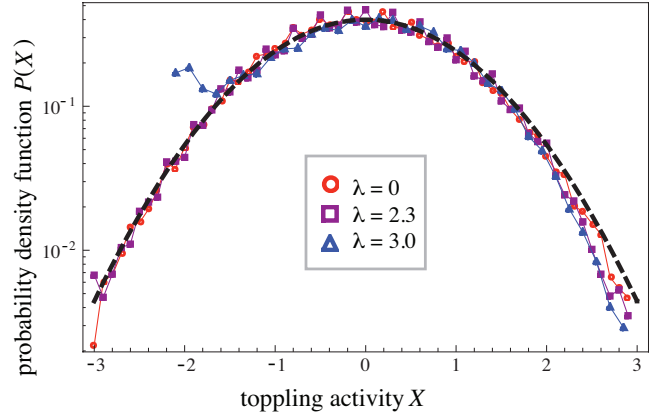}
\caption{{\small The PDF of the fluctuations in the toppling activity of the strongly driven Zhang sandpile driven BTW sandpile
for different values of $\lambda$ (see text). The dashed line is a fitted Gaussian.}} \label{fig8}
\end{center}
\end{figure}
\noindent
\begin{figure}[t]
\begin{center}
\includegraphics[width=8.5cm]{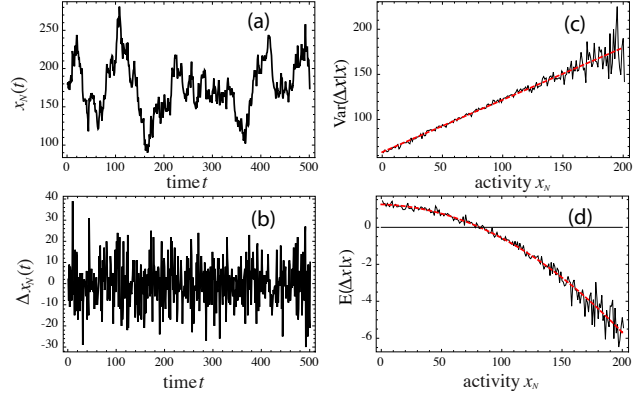}
\caption{{\small a) Part of a time series for the toppling activity in a strongly driven Zhang sandpile. b)
The increments $\Delta x(t)=x(t+\Delta t)-x(t)$ of the signal in (a). As for the slowly driven case. c)
The conditional variance of the increments $\Delta x$ given the value of $x$. d) The conditional mean
of increments. This curve is fitted to a parabola. }} \label{fig9}
\end{center}
\end{figure}

Fig.\ref{fig8} shows the PDFs of the toppling activity in the strongly driven Zhang model where the feeding rate
is a Poisson process with characteristic time scales $\lambda=0$ (feeding one unit $\epsilon$ of mass in every time step), $\lambda=2.3$, and
$\lambda=3.0$. For $\lambda=0$ and $\lambda=2.3$ the sandpile is running, in the meaning that avalanches never terminate.
For $\lambda=3$, the sandpile is no longer running, and we see a deviation from the Gaussian shape in the left
tail of the PDF.

The Gaussian PDFs for strongly driven sandpiles can actually be explained in terms of the stochastic model in
Eq.~(\ref{SDE3}). The idea is that the range of fluctuations is confined by the drift term $f(X)$, which now has
a different shape than in the slowly driven sandpile. Fig.\ref{fig9} shows the shape of both the diffusion
coefficient $D(X)=\sigma^2 X/2$ and the drift term for the strongly driven Zhang model. The diffusion coefficient
has the same form as for the slowly driven sandpile, whereas the drift term is well approximated by a parabola:
$f(X)=-aX^2+bX+c$. As explained in Sec.~\ref{sec1} this drift term confines the fluctuations in toppling activity to a
bounded region around the positive root $X_c$ of $f(X)$.

Due to the higher rate of  random feeding, the memory effects described by the Hurst exponent $H \neq 1/2$ becomes
inessential in the strongly driven sandpile, allowing us to give an approximate description of the time dependent
PDF through the Fokker-Planck equation
\begin{equation}
\frac{\p P}{\p t}=-\frac{\p}{\p X} \Big{(} f(X)\,P \Big{)}+\frac{\sigma^2}{2}\,\frac{\p^2}{\p X^2}(X\,P)\,. \label{FP2}
\end{equation}
Stationary solutions of this equation must satisfy
$$
-\frac{d}{d X} \Big{(} f(X)\,P \Big{)}+\frac{\sigma^2}{2}\,\frac{d^2}{d X^2}(X\,P)=0\,,
$$
and substituting $-aX^2+bX+c$ for $f(X)$, we have the Gaussian solution
$$
P(X)=\frac{1}{\sqrt{2 \pi}\,\Sigma}\,\exp{\left( -\frac{(X-\mu)^2}{2\,\Sigma^2}\right)}\,,
$$
with
$$
\mu=\frac{b}{a}\,\mbox{ and }\,\Sigma=\frac{1}{\sqrt{2}}\,\frac{\sigma}{a}\,.
$$

To further substantiate that the toppling dynamics of sandpiles is fundamentally different from the fluctuating
quantities that give rise to BHP and FTG distributions we apply the above analysis to the fluctuations in total kinetic
energy in a  simulation of  two-dimensional Navier-Stokes turbulence. The
two-dimensional geometry is chosen because of the inverse energy cascade in $k$-space caused by merging of small
vortices and formation of large structures, reminiscent of formation of large avalanches from emerging from
localized random perturbation in sandpile models. In the simulation   energy  is injected via a source term on a
characteristic, small spatial scale throughout the simulation area, and is dissipated as loss through the open
boundary.
\noindent
\begin{figure}[t]
\begin{center}
\includegraphics[width=8.5cm]{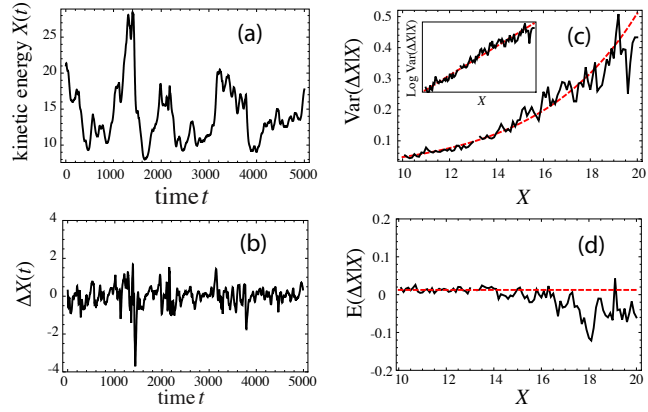}
\caption{{\small a) Part of a time series for the kinetic energy in the 2D turbulence simulation. b) The increments $\Delta X(t)=X(t+\Delta t)-X(t)$ of the signal in (a). As for the toppling activity in sandpiles we
see that the increments are large when the kinetic energy itself is large. c) The conditional variance of
the increments $\Delta X$ given the value of $X$. The inset shows the logarithm of this variance versus $X$. The
dashed curves  corresponds to a fitted exponential function. d) The conditional mean of increments. We
see that the mean is approximately constant for a large range of $X$.}} \label{fig10}
\end{center}
\end{figure}

\begin{figure}[t]
\begin{center}
\includegraphics[width=8.5cm]{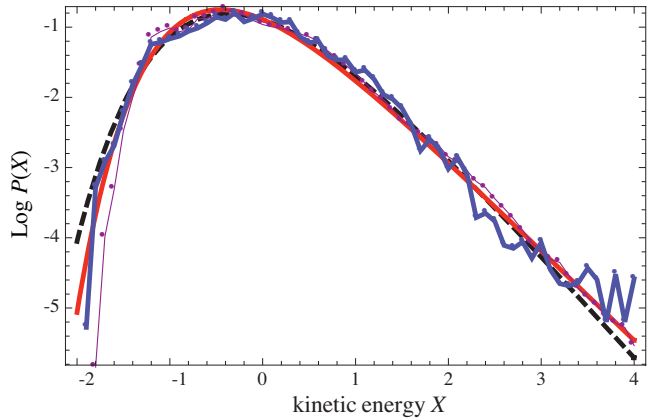}
\caption{{\small The normalized PDF for the fluctuation  of kinetic energy in the 2D turbulence simulation
together with a normalized FTG  PDF (smooth solid curve), the normalized BHP PDF (smooth dashed curve) and a PDF
obtained from the simulation of Eq.~\ref{SDE4} (dotted curve). }} \label{fig11}
\end{center}
\end{figure}

We compute the conditional mean and conditional variance of the increment process. These results are presented in
Fig.\ref{fig10}. We observe that contrary to the sandpile models, the diffusion coefficient for this process grows
exponentially with $X$. Moreover, the drift term can be approximated by a small positive constant except  when the
kinetic energy is very large. This leads us to the stochastic differential equation Eq.~(\ref{SDE4}),
and the corresponding Fokker-Planck equation is
\begin{equation}
\frac{\p P}{\p t}=-b\,\frac{\p P}{\p X}+\frac{c^2}{2}\,\frac{\p^2}{\p X^2} \Big{(} e^{2 a X} P\Big{)}\,. \label{FP3}
\end{equation}
Stationary solutions of this equation must satisfy
$$
-b\,\frac{d P}{d X}+\frac{c^2}{2}\,\frac{d^2}{d X^2} \Big{(} e^{2 a X} P\Big{)}=0\,.
$$
We can put $c=1$ without loss of generality, and obtain the solution
$$
P(X)=\frac{1}{\beta} e^{-(X-\mu)/\beta} e^{-e^{-(X-\mu)/\beta}}\,,
$$
where $\beta=1/2a$ and $\mu=(1/2a) \log{(1/2a) }$. This is the standard FTG distribution. Fig.\ref{fig11} shows
 the normalized PDF for the fluctuations in total kinetic energy obtained from the fluid simulation along with a
normalized FTG distribution and the PDF obtained from the simulation of Eq.~(\ref{SDE4}).
\noindent

\section{Conclusions}\label{sec6}
When output from simple models for avalanching systems are compared to observational data of real-world systems one has to deal with the problem of  establishing a correspondence between the model variables and the observables of the natural system. In general, this is usually not an obvious task, since the model is usually not derived from first physical principles. The observation may be spatiotemporal, or just temporal. Likewise we may choose to analyze the spatiotemporal output from an avalanche model, or just some spatially integrated quantity like the total activity variable in a sandpile, presented as a time series. In this paper we have focused on the latter, and in particular on  reproducing the statistical properties of such time-series by modeling the stochastic process by means of a stochastic differential equation.

Since our interest is  avalanching dynamics  we focus on avalanche statistics, and we have to face the problem of how to define what an avalanche is when our observational data is in the form of a time series. In general we cannot expect that such a definition is necessarily equivalent with a definition based on spatiotemporal data, at least not in those cases where feeding of new ``sand grains" occurs while avalanches are running. In those cases several spatially separated avalanches may run simultaneously, but these cannot be separated in a pure temporal analysis. For such a continuously driven model system the temporal signal may never be zero, and in an observational time series noise also makes it impossible to use a zero signal condition to separate an avalanching state from a quiet state. Thus, the natural way out is to define the avalanching state by means of a threshold level on the activity signal, giving rise to the concept of type-II avalanches.

In this paper we have shown that the toppling activity in sandpiles, and  also global kinetic energy in a 2D fluid simulation, can be modeled by  stochastic differential equations. The modeling clarifies that the main discrepancy between the mean-field approach and the actual BTW and Zhang models is related to the Hurst exponent of the activity process, which is different for the two sandpile models. 
It also clarifies the origin of the differences between the scaling exponents for type-I and type-II avalanches, and why type-II avalanches exhibit clearer scaling characteristics than type-I avalanches.

It follows from the theory how to rescale coordinates to approach the thermodynamic limit, and the results obtained for finite-size scaling in the BTW model in Sec.~\ref{sec4} gives a strong indication that this limit actually exists. 

For continuously driven sandpiles the  stochastic equation can be cast into a Fokker-Planck equation due to the loss of memory caused by the random feeding. This allows an analytic solution for the activity PDF, which is a Gaussian for the sandpile models, but gives rise to the FTG distribution for the 2D fluid simulation. These results show that the non-Gaussian universal PDF described in \cite{Bramwell1,Bramwell2} is not relevant for strongly driven sandpiles, but may be so for certain  turbulence models. The stochastic theory relates the difference between Gaussian and FTG distributed activity signal to a the difference between a linear and exponential $X$ dependence of the diffusion coefficient in the stochastic equation.

\end{document}